\documentclass[prl,twocolumn,unsortedaddress,showpacs,floatfix]{revtex4}

\usepackage{graphicx}
\usepackage{array}
\usepackage{amsmath,amsfonts,amssymb}

\begin{document}

\title{Atom Transfer and Single-Adatom Contacts}

\author{L.  Limot}
\author{J.  Kr\"{o}ger}
\author{R.  Berndt}
\affiliation{Institut f\"{u}r Experimentelle und Angewandte
Physik, Christian-Albrechts-Universit\"{a}t zu Kiel, D-24098 Kiel,
Germany}

\author{A.  Garcia-Lekue}
\author{W.  A.  Hofer}
\affiliation{Surface Science Research Centre, University of
Liverpool, Liverpool L69 3BX, United Kingdom}

\date{\today}

\begin{abstract}
The point contact of a tunnel tip approaching towards Ag(111) and
Cu(111) surfaces is investigated with a low temperature scanning
tunneling microscope. A sharp jump-to-contact, random in nature,
is observed in the conductance. After point contact, the tip-apex
atom is transferred to the surface, indicating that a one-atom
contact is formed during the approach. In sharp contrast, the
conductance over single silver and copper adatoms exhibits a
smooth and reproducible transition from tunneling to contact
regime. Numerical simulations show that this is a consequence of
the additional dipolar bonding between the homoepitaxial adatom
and the surface atoms.
\end{abstract}


\maketitle How do mechanical and transport properties change as
matter is sized down to the atomic scale? This question is
considered of fundamental interest and potentially important in
view of future nanoscale device technologies. It is also of
particular interest for our understanding of technologically
important problems like friction, machining, lubrication and
adhesion, where the contact between macroscopic bodies occurs
typically at numerous atomic-size protrusions, whose properties
determine those of the macroscopic contact. Proximity probes like
the Scanning Tunneling Microscope (STM), metal break junctions and
related techniques, together with computational methods for
simulating tip-sample interactions with atomic detail, have
enabled to address this question by investigating atomic-size
contacts \cite{agr03}. Metallic point contacts between two
metallic surfaces are known to exhibit a jump in the conductance
\cite{gim87,dur90,ole96}. When stretched to the point of breaking,
their conductance decreases in discrete steps of $\sim 2e^2/h$
\cite{pas93,kra93,ole94}, the conductance expected for a
one-dimensional conductor with one propagating channel. Since the
shape of this staircase is material-dependent, the chemical
valance of the contact-atoms is likely fixing the number of
conduction channels \cite{sch98}.
\newline\- The reverse process -- a STM tip approaching a metallic surface at
close range and creating a contact, as opposed to the stretching
and breaking of contacts -- is far less documented. This is quite
surprising since the STM allows for measurements with knowledge of
identity, location and number of atoms in between electrodes. A
pioneering investigation of this type was performed on a single Xe
atom and a two-Xe-atom chain using a W tip and a nickel surface
\cite{azd96}. Xe did not exhibit metallic behavior, which is
reflected by the low conductance observed for the contacts (below
$0.2$ in units of $2e^2/h$). In the present Letter, we report an
investigation of well defined metallic contacts. Using a low
temperature STM, we create single-atom contacts by lowering the
tip over isolated metallic adatoms, Ag atoms on Ag(111) and Cu
atoms on Cu(111). Unlike tip-surface contacts where a sharp jump
in the conductance is observed, tip-adatom contacts exhibit a
smooth and reproducible variation from tunneling to contact
regime. A numerical analysis indicates that the additional dipolar
bonding of the adatom compared to surface atoms explains this
surprising finding. Moreover, we show that the jump-to-contact
over the clean surfaces also results in the transfer of the
tip-apex atom to the surface. This is experimental evidence that a
one-atom contact is formed when the tip is in close proximity of a
clean surface, as usually inferred from the value of $G\sim 1$ (in
units of $2e^2/h$) at point contact.
\newline\- The experiments were performed in a custom-built ultrahigh vacuum
STM operating at $4.6$ K using Ag(111) and Cu(111) surfaces --
cleaned by Ar$^{+}$ sputter/anneal cycles. The W tip was first
electrochemically etched \textit{ex situ}, and then prepared
\textit{in situ} by soft indentations into the surface, until
adatoms were imaged spherically (see Fig.~\ref{fig2}c). Given this
preparation, the tip is covered with surface material. The
conductance ($G=I/V$, where $I$ is the tunneling current) versus
tip excursion ($\Delta z$) was measured by opening the feedback
loop at $V=100$ mV (200 mV) in the center of defect- and
impurity-free areas ($20 \times 20$ nm$^{2}$) of Ag (of Cu). The
tip was then driven towards the surface at rates ranging from 1 to
2 {\AA}/s recording concomitantly the conductance, and finally
retracted by $40$ {\AA}. The experimental setup of
Ref.~\onlinecite{lim03} was employed to ensure that the high
conductance measurements were not polluted by a voltage drop at
the input impedance of the current preamplifier.
\begin{figure}[t]
\includegraphics[width=5.0cm,bbllx=10,bblly=70,bburx=560,bbury=795,clip=]{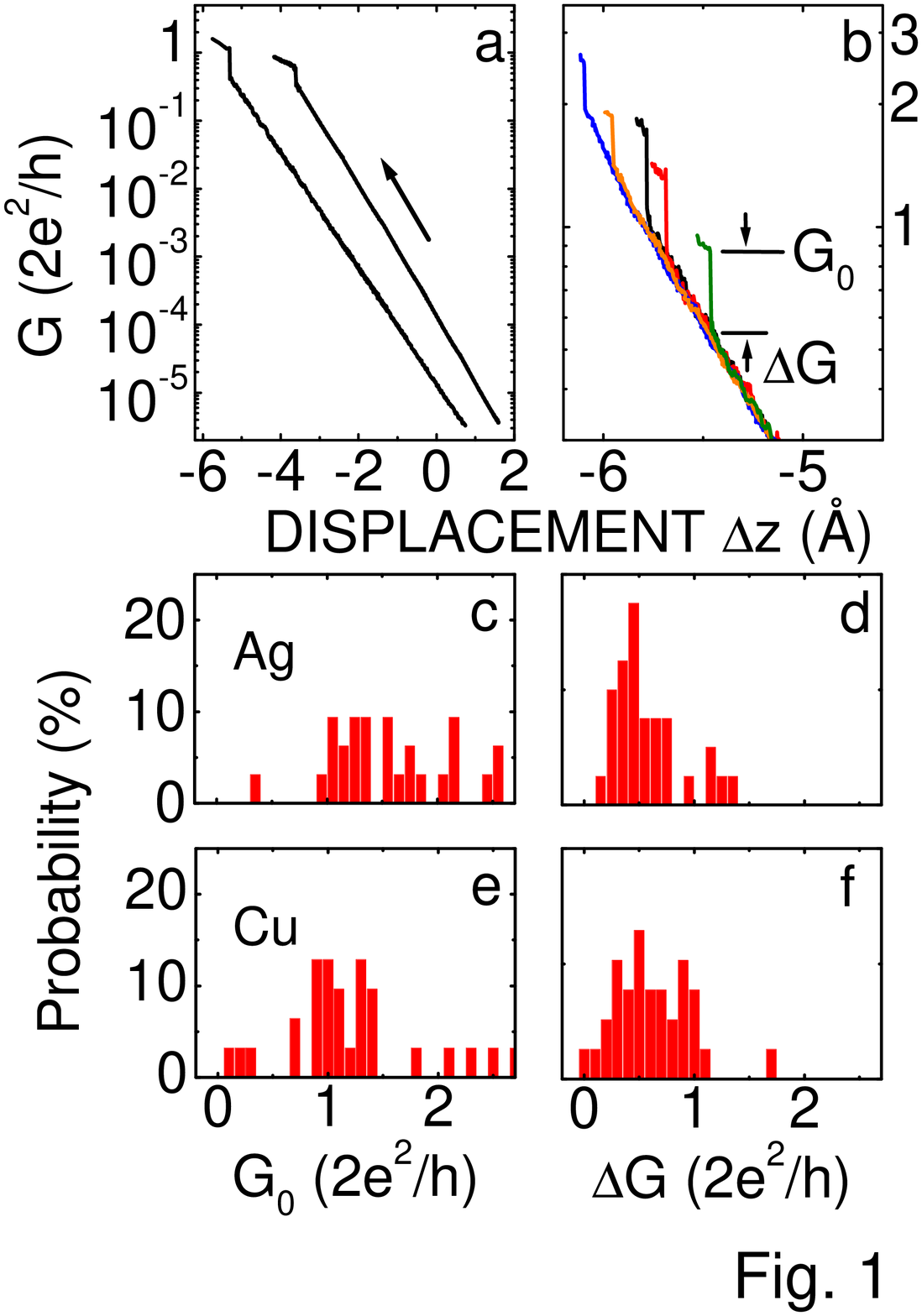}
\caption{a) Experimental $G$ in units $2e^{2}/h$ versus $\Delta z$
for Ag(111) (left) and Cu(111) (right) -- arrow indicates
direction of tip motion. The origin $\Delta z=0$ is arbitrarily
fixed at $G=1$ nS for Ag(111) and $G=10$ nS for Cu(111). b) Degree
of reproducibility of the point contact transition for Ag(111),
and histograms of c) the conductance $G_0$ after point contact, d)
the conductance-jump $\Delta G$. Histograms of e) $G_0$, f)
$\Delta G$ for Cu(111).}\label{fig1}
\end{figure}
\newline\- A typical $G$ versus $\Delta z$ measurement over Ag(111) and
Cu(111) is presented in Fig.~\ref{fig1}a. As the tip is approached
towards the surface, the conductance increases exponentially up to
$G\sim 1$ where a sudden jump-to-contact is observed. From the
exponential tunneling behavior, where $G \propto
\text{exp}(-1.025\sqrt{\phi}\ \Delta z)$, we extract an apparent
barrier height $\phi=4.0(2)$ eV and $\phi=4.7(2)$ eV for Ag(111)
and Cu(111) respectively, typical of metals \cite{ole96}. A series
of jumps acquired with different tips and at different locations
of the Ag(111) surface are presented in Fig.~\ref{fig1}b. As
shown, the conductance after the jump ($G_0$), the height of the
jump ($\Delta G$) and the tip excursion at which the jump occurs
are not completely reproducible. A survey carried out over 30
jumps observed over Ag(111) (Figs.~\ref{fig1}c and ~\ref{fig1}d)
indicates that $G_0=1.5$ and $\Delta G=0.4$, with large standard
deviations of 0.6 and 0.2, respectively. Similarly, a survey over
Cu(111) yields $G_0=1.1(3)$ and $\Delta G=0.6(4)$
(Figs.~\ref{fig1}e and ~\ref{fig1}f).
\newline\- Figure~\ref{fig2} illustrates an important experimental
finding of this Letter. Along with the conductance measurements
presented in Fig.~\ref{fig1}, we have acquired images prior to
(Fig.~\ref{fig2}a) and after (Fig.~\ref{fig2}b) the observation of
a conductance jump. After a jump, our measurements indicate that
there is a $75$\% probability for Ag ($60$\% for Cu) that a single
atom remains on the surface at the location where the $G$ vs
$\Delta z$ was performed (marked by an arrow in Fig.~\ref{fig2}a),
and a nearly $25$\% probability ($40$\% for Cu) to find a cluster
of atoms (Fig.~\ref{fig2}c) -- in $1$ out of $25$ approaches no
material is transferred to the surface. Figure~\ref{fig2}c shows
that atoms can be transferred in this way from the tip to a
desired location of the surface with sub-nanometric precision.
Given our \textit{in situ} tip preparation, the deposited atoms
are homoepitaxial atoms -- Ag atoms for Ag(111) and Cu atoms for
Cu(111). To confirm this, single atoms were extracted from the Ag
and the Cu substrates by indenting the tip $10$ {\AA} into the
surface. These atoms yielded the same profile in the STM images
and the same ${\rm d}I/{\rm d}V$ spectrum \cite{lim04} as the
atoms deposited from the tip. This is also an indication that the
atom transfer from tip-to-surface does not damage the substrate in
the point-contact region.
\begin{figure}[t]
\includegraphics[width=5.0cm,bbllx=65pts,bblly=365,bburx=530,bbury=775,clip=]{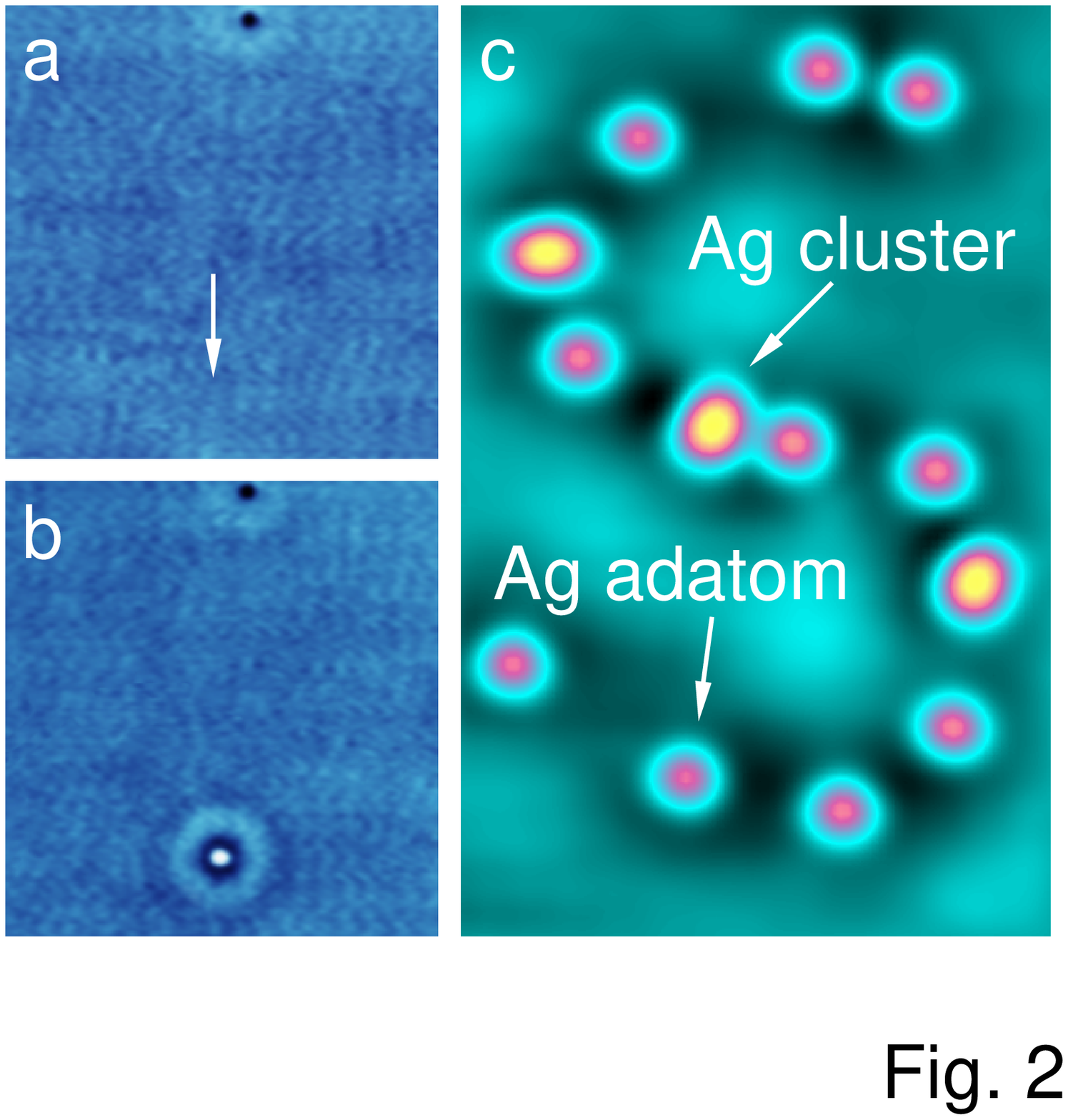}
\caption{Constant current images of the Ag(111) surface ($24
\times 24$ nm$^{2}$, 100 mV, 0.1 nA): a) Prior to, b) After the
point-contact transition in the conductance (arrow indicates where
the tip was approached). c) Constant current image acquired after
a series of point-contact transitions ($8 \times 12$ nm$^{2}$, 100
mV, 0.1 nA).} \label{fig2}
\end{figure}
\newline\- Figure~\ref{fig3} presents another important finding.
To obtain fully reproducible experimental data we extended our
study to the conductance of individual homoepitaxial adatoms. A
typical adatom conductance versus displacement curve is shown in
Fig.~\ref{fig3}a. The exponential behavior expected for tunneling
is observed up to $G\approx 0.1$, but with higher apparent barrier
height of $\phi=4.6(2)$ eV for Ag and of $\phi=5.3(2)$ eV for Cu
compared to the clean surfaces. Most importantly, the adatom
conductance does not exhibit a sharp jump. Rather, a
``glide-to-contact'' is observed: a smooth upturn with respect to
the tunneling behavior which sets in at $G\approx 0.3$ and extends
over $\approx 0.5$ {\AA}, followed by a saturation at a
conductance $G_0$, defined here by the intersection between the
experimental curve and the extrapolated tunneling conductance at
large tip-sample separations (see Fig.~\ref{fig3}a). No material
is transferred to the surface. A survey carried out with 20
different tips (on average, 10 approaches were performed for each
tip) indicates that $G_0$ is highly reproducible at the adatoms,
contrary to the scatter observed on the clean surfaces, with a
mean value of $G_0=0.93(5)$ for Ag (Fig.~\ref{fig3}b), and of
$G_0=0.98(6)$ for Cu (Fig.~\ref{fig3}c), in agreement with
break-junctions studies of noble-metal contacts \cite{agr03}.
\begin{figure}[t]
\includegraphics[width=5.0cm,bbllx=20,bblly=50,bburx=525,bbury=760,clip=]{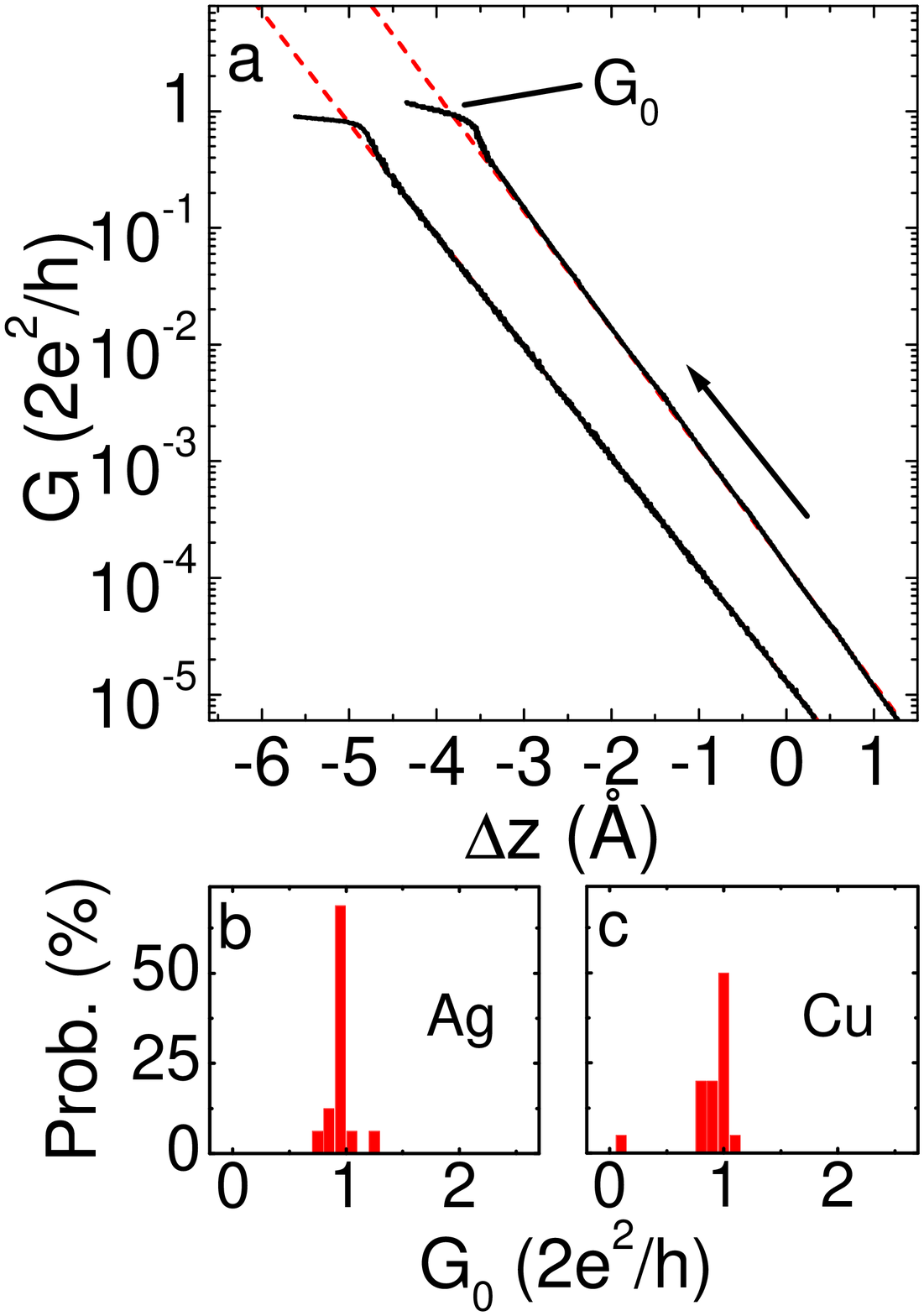}
\caption{a) Experimental $G$ versus $\Delta z$ (solid line) and
extrapolated tunneling conductance (dashed line) for Ag/Ag(111)
(left) and for Cu/Cu(111) (right). The arrow indicates the
direction of tip motion ($\Delta z=0$ at $G=1$ nS for Ag and at
$G=10$ nS for Cu). b) Ag/Ag(111), and c) Cu/Cu(111) histograms of
the conductance $G_0$ after point contact (same binning as in
Figs.~\ref{fig1}c and ~\ref{fig1}e).} \label{fig3}
\end{figure}
\newline\- Before discussing the adatom case, we first focus on the
conductance measurements over the clean surfaces. A number of
authors have modeled the relaxations effects in a STM when the tip
is approached in close proximity of a metallic surface
\cite{dur90,ole96,hof01}. It was shown that at interatomic
distances of order the typical bond length of the atoms, both tip
and surface stretch towards each other because of attractive
adhesive forces acting between them. Consistent with the findings
of Olsen \textit{et al.} \cite{ole96}, we observe, within
experimental error, an exponential variation of $G$ with $\Delta
z$ down to point contact, which, as Olsen \textit{et al.} pointed
out, is the signature of relaxations effects setting in at least
$1$ {\AA} before the jump-to-contact occurs. Another experimental
signature of relaxation effects is the enhanced Stark shift
observed in this range for the ${\rm d}I/{\rm d}V$ spectra of the
Ag(111) surface state \cite{lim03}.
\newline\- The jump-to-contact occurs when chemical bonds between
the surface and the tip apex start to weaken the adhesion of the
atom to the tip structure. In this case, and over a relatively
small distance variation of less than $0.1$ {\AA} \cite{hof04},
the atom will be transferred from the tip onto the surface. In
principle, such an atomic transfer could be reversible. However,
this was not observed experimentally. To understand the origin of
this behavior in detail, we modeled a coupled Cu system by a flat
(3 $\times$ 3) unit cell and a tip consisting of a Cu pyramid
mounted on the reverse of the five layer surface film
\cite{hof01}. At a core-core distance of $3$ {\AA} between tip
apex and surface atoms we performed three separate sets of
calculations by standard density functional methods
\cite{kre93kre96}: (i) The apex atom was transferred from the tip
onto the surface, while all other atoms were kept frozen. (ii) The
tip pyramid and the surface layers were fully relaxed. (iii) The
tip pyramid was kept frozen, only surface atoms were relaxed. In
the first case we observe a parabolic energy distribution, the
minimum energy corresponds to a median distance of the apex atom
from both surfaces. In the second case we observe strong outward
relaxation of the surface layer, coupled to the transfer of the
apex atom onto the surface. The total energy gain by relaxations
amounts to about $1$ eV/atom in the interface. In the third case
our numerical methods did not arrive at a stable solution. We
conclude from these results that the tip apex atom will be
transferred in every case, once the distance is well below the
jump-to-contact point. Under ambient thermal conditions
transferred tip atoms may diffuse rapidly to the step edges of the
crystal and therefore not be observed. The analysis also confirms
that a one-atom contact is formed during the approach, which is
otherwise inferred from the value of $G$ close to the quantum of
conductance.
\newline\- The random character of the jump-to-contact
over the clean surfaces (Fig.~\ref{fig1}) can be understood in the
light of recent simulations \cite{hof01}. It was shown that the
jump-to-contact strongly depends on where the approach is
performed on the surface. Regardless of its chemical nature, when
the tip is positioned on top of a surface atom, the jump should be
detected about $0.5$ {\AA} earlier compared to a three-fold hollow
position, all other locations on the surface exhibiting a jump
within this range. Following this viewpoint, since surface atoms
are not usually resolved  -- this is quite common for compact
(111) surfaces -- the conductance measurements are performed at
random locations of the surface, and the jumps occurs randomly
within a finite excursion range of the tip. Based on the data,
this range is estimated to be $0.5$ {\AA} for Ag
(Fig.~\ref{fig1}b) and $0.3$ {\AA} for Cu, in good agreement with
the above prediction. The random nature of $G_0$ and of $\Delta G$
indicate that conformation changes of the interface may affect
electronic transport properties quite substantially, as recent
transport simulations on Al(111) pointed out \cite{blan04}. Over
the adatoms, however, where the two systems approach contact on a
well defined location, the conductance is reproducible
(Figs.~\ref{fig3}b and \ref{fig3}c).
\newline\- The atomic arrangement and the conductance properties
of single adatoms Cu/Cu(111) and Ag/Ag(111) was simulated by two
sets of calculations: (i) In a density functional simulation of a
five layer surface film with an atomically sharp tip model we
calculated the relaxed positions of surface and tip at intervals
of 0.2 {\AA}. (ii) The conductance properties were also calculated
during the approach, taking into account the shift of atomic
positions with a recently developed model \cite{hof04}. As the
first set of simulations reveals, the atomic positions are only
slightly relaxed during the transition from the tunneling regime
to point contact (Fig.~\ref{fig4}a). This is in sharp contrast to
the situation on a flat surface, where the tip will be fractured,
as shown above by experiments and simulations (see also
Fig.~\ref{fig4}b). The reason for this marked difference is the
larger stiffness of the adatom. We find an increase of the elastic
constants  on Cu (Ag) to $6.1$ $(5.1)$ eV/\AA$^2$, which is nearly
double the value found on flat surfaces \cite{hof04}. This
indicates that the redistribution of surface charge due to the
Smoluchowski effect creates a surface dipole which enhances the
bonding of the adatom. The larger stiffness of the adatom bond is
joined by a lower interaction energy with the tip. In both cases
it remains well below 1 eV, which was identified as the threshold
value for jump-to-contact and atom transfer (Fig.~\ref{fig4}b).
Based on these findings, it can be concluded that the reproducible
transition from the tunneling to the contact regime is due to the
comparatively small and reversible relaxations.
\begin{figure}[t]
\includegraphics[width=6.8cm,bbllx=10,bblly=330,bburx=565,bbury=750,clip=]{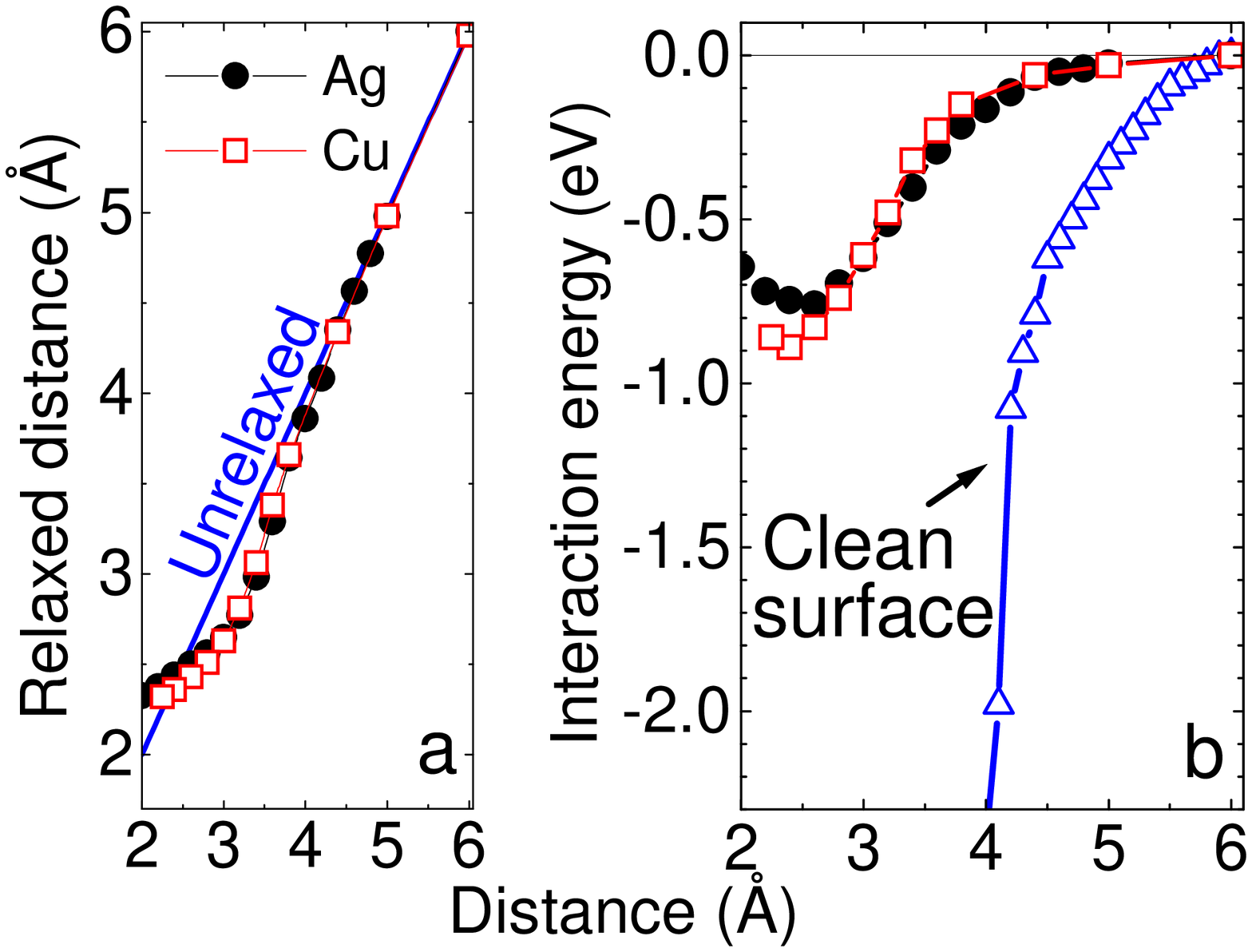}
\caption{Calculated results for adatoms. a) Actual tip-adatom
distance (relaxed distance) versus distance in the absence of
tip-sample interaction. Intersection with the unrelaxed curve
(solid line) indicates transition from attractive to repulsive
regime. b) Adatom interaction with the tip versus distance. The
onset of strong chemical attractions occurs at about $4$ {\AA}.
Open triangles: calculated tip interaction with the clean Au(111)
surface \cite{hof01,hof04}. Tip fracture occurs at $4.2$ {\AA}.}
\label{fig4}
\end{figure}
\newline\- The conductance was simulated with Bardeen's method. The only effect
of interactions included in the simulation was the relaxation of
the adatom. However, in this situation the strong localization of
the adatom wave functions, in combination with an atomically sharp
tip, will lead to large changes of the crystal potential in the
immediate vicinity of the point contact even for larger distances
\cite{stok99}. Since the exponential decay of the wave functions
reflects these potential changes, perturbation methods become
unreliable. The exponential decay constants were therefore
calculated for large distances of 9 {\AA}, where the two systems
will be completely decoupled. Here, at the position of the adatom
(on the flat surface) the apparent barrier height is $\phi=5.0$
$(4.4)$ eV for Cu and $\phi=4.6$ $(4.0)$ eV for Ag. The simulated
values agree very well with experimental findings. Interestingly,
a drop of the potential barrier by about $1$ eV at a distance of
$6$ {\AA}, is also observed in the simulations. That the barrier
height remains close to constant in this range cannot be due to
atomic relaxations. It seems, therefore, that the localization of
electron states on single atoms at both sides of the tunneling
junction leads to a substantial change of the potential barrier
even in this range. We shall analyze this feature in future
simulations.
\newline\- Summarizing, a sharp jump in the conductance is observed
when a metallic tip is brought into contact with a compact (111)
metallic surface. The jump is associated with an irreversible tip
fracture, which, in most cases, results in the transfer of the
tip-apex atom to the surface. This is experimental evidence that a
one-atom contact is formed. When contacting single adatoms, on the
contrary, no material is transferred, and a smooth and
reproducible transition occurs from tunneling to contact regime.
Single adatom-contacts with a STM are therefore junctions which
can potentially be employed to probe electron transport through
single atoms, and eventually molecules, with knowledge of number
and identity of the atoms in between electrodes.
\newline\- We gratefully acknowledge discussions with P.\ Johansson.
L.\ L., J.\ K. and R.\ B. thank the Deutsche
Forschungsgemeinschaft for financial support. W.\ A.\ H. thanks
the Royal Society for the award of a University Research
Fellowship. A.\ G.-L. is funded by a Marie Curie Fellowship from
the European Commission.


\begin{thebibliography}{99}

\bibitem{agr03} N.\ Agra\"{\i}t, A.\ L.\ Yeyati, and
J.\ M.\ van Ruitenbeek, Physics Rep. \textbf{377}, 81 (2003).

\bibitem{gim87} J.\ K.\ Gimzewski and R. M\"{o}ller, \prb
\textbf{36}, R1284 (1987);

\bibitem{dur90} U.\ D\"{u}rig, O.\ Z\"{u}ger, and D.\ W.\ Pohl, \prl \textbf{65},
349 (1990).

\bibitem{ole96} L.\ Olesen, M.\ Brandbyge, M.\ R.\ S{\o}rensen, K.\ W.\ Jacobsen,
E. L{\ae}gsgaard, I.\ Stensgaard, and F.\ Besenbacher, \prl
\textbf{76}, 1485 (1996).

\bibitem{pas93} J.\ I.\ Pascual, J.\ M\'{e}ndez, J.\ G\'{o}mez-Herrero,
A.\ M.\ Bar\'{o}, N.\ Garc\'{\i}a, and V.\ T.\ Binh, \prl
\textbf{71}, 1852 (1993).

\bibitem{kra93} J.\ M.\ Krans, C.\ J.\ Muller, I.\ K.\ Yanson, Th.\ C.\ M.\ Govaert,
R.\ Hesper, and J.\ M.\ van Ruitenbeek, \prb \textbf{48}, R14721
(1993).

\bibitem{ole94} L.\ Olesen, E.\ L{\ae}gsgaard, I.\ Stensgaard, F.\ Besenbacher, J.\ Schi{\o}tz,
P.\ Stoltze, K.\ W.\ Jacobsen, and J.\ K.\ N{\o}rskov, \prl
\textbf{72}, 2251 (1994).

\bibitem{sch98} E.\ Scheer, N.\ Agra\"{\i}t, J.\ C.\ Cuevas, A.\ L.\ Yeyati, B.\ Ludoph,
A.\ Martin-Rodero, G.\ R.\ Bollinger, J.\ M.\ van Ruitenbeek, and
C.\ Urbina, Nature (London) \textbf{394}, 154 (1998).

\bibitem{azd96} A.\ Yazdani, D.\ M.\ Eigler, and N.\ D.\ Lang,
Science \textbf{272}, 1921 (1996).

\bibitem{lim03} L.\ Limot, T.\ Maroutian, P.\ Johansson, and R.\ Berndt, \prl
\textbf{91}, 196801 (2003).

\bibitem{lim04} L.\ Limot, E.\ Pehlke, J.\ Kr\"{o}ger, and R.\
Berndt, CondMat/0405397.

\bibitem{hof01} W.\ A.\ Hofer, A.\ J.\ Fisher, R.\ A.\ Wolkow, and P.\ Gr\"{u}tter, \prl \textbf{87},
236104 (2001).

\bibitem{hof04} W.\ A.\ Hofer, A.\ Garcia-Lekue, and H.\ Brune, Chem. Phys. Lett. \textbf{397},
354 (2004).

\bibitem{kre93kre96} G.\ Kresse and J.\ Hafner, \prb \textbf{47}, R558
(1993); G.\ Kresse and J.\ Furthm\"{u}ller, \textit{ibid.}
\textbf{54}, 11169 (1996).

\bibitem{blan04} J.\ M.\ Blanco, C.\ Gonzalez, P.\ Jelinek, J.\ Ortega, F.\ Flores, and
R.\ Perez, \prb \textbf{70}, 085405 (2004).

\bibitem{stok99} K.\ Stokbro, Surf. Sci. \textbf{429}, 327 (1999).

\end{thebibliography}
\end{document}